\documentclass[lettersize,journal]{IEEEtran}

\usepackage{amsmath,amsfonts}
\usepackage{algorithmic}
\usepackage{algorithm}
\usepackage{array}
\usepackage{textcomp}
\usepackage{stfloats}
\usepackage{url}
\usepackage{verbatim}
\usepackage{graphicx}
\usepackage{cite}

\usepackage{booktabs}
\usepackage{adjustbox}
\usepackage{caption}
\usepackage{subcaption}
\usepackage{color,soul}
\usepackage{multirow}
\usepackage{siunitx}

\begin{document}

\title{Missing Data Estimation for MR Spectroscopic Imaging via Mask-Free Deep Learning Methods}
\author{Tan-Hanh Pham, Ovidiu C. Andronesi, Xianqi Li, and Kim-Doang Nguyen
\thanks{Tan-Hanh Pham is a research fellow at Florida Institute of Technology and an oncoming research fellow at Massachusetts General Hospital, Harvard Medical School, USA (e-mail: hanhpt.phamtan@gmail.com).}
\thanks{Ovidiu C. Andronesi is with the A. A. Martinos Center for Biomedical Imaging, Department of Radiology, Massachusetts General Hospital, Harvard Medical School, USA (e-mail: OANDRONESI@mgh.harvard.edu).}%
\thanks{Xianqi Li is with the Department of Mathematics and Systems Engineering, Florida Institute of Technology, USA (e-mail: xli@fit.edu).}%
\thanks{Kim-Doang Nguyen is with the Department of Mechanical and Civil Engineering, Florida Institute of Technology, USA (e-mail: knguyen@fit.edu).}%
\thanks{\textit{Corresponding authors}: Xianqi Li; Kim-Doang Nguyen.}
}



\maketitle

\begin{abstract}
Magnetic Resonance Spectroscopic Imaging (MRSI) is a powerful tool for non-invasive mapping of brain metabolites, providing critical insights into neurological conditions. However, its utility is often limited by missing or corrupted data due to motion artifacts, magnetic field inhomogeneities, or failed spectral fitting-especially in high resolution 3D acquisitions. To address this, we propose the first deep learning-based, mask-free framework for estimating missing data in MRSI metabolic maps. Unlike conventional restoration methods that rely on explicit masks to identify missing regions, our approach implicitly detects and estimates these areas using contextual spatial features through 2D and 3D U-Net architectures. We also introduce a progressive training strategy to enhance robustness under varying levels of data degradation. Our method is evaluated on both simulated and real patient datasets and consistently outperforms traditional interpolation techniques such as cubic and linear interpolation. The 2D model achieves an MSE of 0.002 and an SSIM of 0.97 with 20\% missing voxels, while the 3D model reaches an MSE of 0.001 and an SSIM of 0.98 with 15\% missing voxels. Qualitative results show improved fidelity in estimating missing data, particularly in metabolically heterogeneous regions and ventricular regions. Importantly, our model generalizes well to real-world datasets without requiring retraining or mask input. These findings demonstrate the effectiveness and broad applicability of mask-free deep learning for MRSI restoration, with strong potential for clinical and research integration.

\end{abstract}


\section{Introduction}
\IEEEPARstart{M}{agnetic} Resonance Spectroscopy Imaging (MRSI) is a powerful non-invasive imaging technique that enables the measurement of tissue metabolites, offering unique insights into brain metabolism \cite{bogner2021accelerated,brown1982nmr} beyond what is accesible through structural MRI. By detecting metabolites such as N-acetylaspartate (NAA), creatine, choline, and lactate \cite{maudsley1983spatially,sabati2015multivendor}, MRSI provides crucial information for diagnosing and monitoring neurological disorders, including gliomas, epilepsy, and neurodegenerative diseases \cite{li2017advanced,li2020super}. These metabolic signatures can reveal early-stage biochemical changes, making MRSI an essential tool for research and clinical decision-making.

However, MRSI data acquisition is often compromised by several challenges, primarily due to artifacts caused by patient motion, magnetic field inhomogeneities, and technical limitations that decrease data quality \cite{andronesi2021motion,juchem2021b0,tkavc2021water}. These issues are especially pronounced in high-resolution 3D acquisitions, where long scan times and complex voxel-wise spectral processing introduce variability in data quality. As a result, a significant portion of the MRSI voxels may be rendered unusable, degrading the reliability of downstream analysis and visualization. Conventional solutions typically rely on interpolation techniques (e.g., linear, cubic, spline), which are easy to implement but often insufficient for preserving structural and pathological features, especially in metabolically heterogeneous regions like tumor margins \cite{lam2016high,klauser2019fast,wu2022slice}. More recently, inpainting approaches \cite{garcia2010robust,torrado2021inpainting} have been explored to estimate missing voxels, showing performance improvements over interpolation. However, these methods require explicit binary masks to indicate missing regions, a major limitation in real-world settings where mask generation is non-trivial or unavailable.

Recent advances in deep learning have significantly improved image segmentation, reconstruction, and inpainting in medical imaging \cite{pham2024seunet,reader2020deep,armanious2019adversarial}. For example, encoder-decoder networks like U-Net \cite{ronneberger2015u} and their 3D variants \cite{cciccek20163d,hatamizadeh2021transformers} have achieved strong performance in tasks such as segmentation, denoising, and artifact removal. Yet, in the domain of MRSI, deep learning remains largely underutilized. Most prior efforts have focused on spectral quality control \cite{dziadosz2023denoising,van2023review,rakic2024deep} and super-resolution \cite{iqbal2019super,li2021nimg,li2022deep,li5228203super}, rather than on the voxel-wise recovery of missing metabolic data. In this study, we propose a novel, deep learning-based framework for reconstructing missing voxels in MRSI metabolic maps. To our knowledge, this is the first work to apply deep learning to this problem using a mask-free strategy. Instead of relying on predefined masks to locate missing voxels, our method learns to identify and restore them implicitly using only spatial context from the input data. We implement both 2D and 3D U-Net based models and introduce a progressive training strategy that gradually increases the difficulty of missing data scenarios. This strategy enhances model robustness and generalization to real-world cases with varied missing patterns.
Our key contributions include:
\begin{itemize}
    \item A mask-free deep learning framework for reconstructing missing voxels in MRSI metabolic maps, eliminating the need for explicit region-of-interest masking. 
    \item Incorporating a progressive training strategy that gradually exposes the model to increasing levels of data degradation, enhancing its robustness;
    \item Validation on both synthetic and real patient datasets, demonstrating strong generalization and structural fidelity;
    \item Outperforming traditional interpolation methods in quantitative metrics (MSE, SSIM) and in visual data missing estimation quality - especially in regions with pathological significance.
\end{itemize}

The rest of the paper is organized as follows: Section \ref{sec.materialandmethod} presents the dataset processing used in the study and the proposed method. Section \ref{sec.experimentsetup} describes the experimental settings and evaluation metrics. The results and discussions are presented in section \ref{sec.result}. Finally, section \ref{sec.discussion} summarizes the key findings, provides conclusion remarks, and suggests potential directions for future research.

\section{Material and Methodology}
\label{sec.materialandmethod}
This section outlines the complete pipeline developed to address the problem of missing voxel estimation in MRSI metabolic maps. The methodology is composed of: (1) Synthetic data generation simulating realistic MRSI volumes, (2) Simulation of missing voxels due to motion artifacts, magnetic field inhomogeneities, or spectral fitting failures, and (3) A supervised-learning estimation for data missing framework based on 2D and 3D U-Net architectures, trained with a progressive strategy to improve robustness.

\subsection{Generation of Synthetic MRSI Volumes}

Due to the limited availability of large-scale, high-resolution (HR) MRSI datasets with ground truth, we generate synthetic MRSI volumes from multi-modal MRI scans of 75 patients. FLAIR images are used to segment tumor regions using ITK-SNAP \cite{yushkevich2006user}, while T1-weighted MEMPRAGE images are segmented into gray matter (GM), white matter (WM), and cerebrospinal fluid (CSF) using FSL tools \cite{jenkinson2012fsl}. The synthetic HR MRSI maps are created using a weighted sum:
\begin{equation}
\mbox{MRSI}_{HR} = 0.1 \times \text{GM} + 0.12 \times \text{WM} + 0 \times \text{CSF} + \tau \times \text{TM},
\end{equation}
where the tumor weighting factor $\tau$ is sampled from $[0.2, 0.3,..., 0.8]$ to simulate a range of pathological metabolite intensities. The resulting maps are normalized to $[0,1]$ and downsampled to $64 \times 64 \times 64$ volumes, forming the input for downstream model training. In total, we have 75 3D samples, which are divided into two subsets: 60 samples for training and 15 samples for testing.

\subsection{Simulation of Missing Voxels}
\label{sec.data_processing}
To train and evaluate the proposed framework under realistic conditions, we simulate missing voxels in the synthetic MRSI volumes. These missing voxels are intended to mimic common real-world causes of data corruption in MRSI, such as motion artifacts, field inhomogeneities, and failed spectral fitting. Rather than using arbitrary or uniform dropout patterns, we design a structured simulation protocol that introduces controlled, localized voxel loss within the brain region of each volume, which is described in Algorithm \ref{algorithm_blackingout}.

\begin{algorithm}[ht]
    \caption{Pseudo-code for generating 3D MRSI with filtered voxels.}
    \label{algorithm_blackingout}
    \setlength{\baselineskip}{1.1\baselineskip}
	\begin{algorithmic}[1]
        \STATE  \textbf{Input:} 3D brain volume $\mathbf{V} \in \mathbb{R}^{D \times H \times W}$, noise percentage $\eta$, minimum cluster size $s_{\text{min}}$, maximum cluster size $s_{\text{max}}$, noise type $T$
        \STATE  \textbf{Output:} filtered 3D MRSI brain volume $\mathbf{\hat{V}}$
        \STATE  Identify the brain region by finding non-black voxels $\mathbf{M}$ as:
        \[
        \mathbf{M}(x, y, z) =
        \begin{cases} 
        1, & \text{if } \mathbf{V}(x, y, z) > 0, \\
        0, & \text{otherwise.}
        \end{cases}
        \]
        \STATE Extract bounding box coordinates $\{(z_{\text{min}}, z_{\text{max}}), (y_{\text{min}}, y_{\text{max}}), (x_{\text{min}}, x_{\text{max}})\}$ of $\mathbf{M}$.
        \STATE Identify center of the brain region $\mathbf{c} = \left(\frac{x_{\text{min}} + x_{\text{max}}}{2}, \frac{y_{\text{min}} + y_{\text{max}}}{2}, \frac{z_{\text{min}} + z_{\text{max}}}{2}\right)$.
        \STATE  Compute radius $r = \min\left(\frac{x_{\text{max}} - x_{\text{min}}}{2}, \frac{y_{\text{max}} - y_{\text{min}}}{2}, \frac{z_{\text{max}} - z_{\text{min}}}{2}\right)$.

        \STATE Total brain voxels $N_{\text{brain}} = \sum_{x, y, z} \mathbf{M}(x, y, z)$.
        \STATE Noise voxels to replace $N_{\text{noise}} = \lfloor \eta \cdot N_{\text{brain}} \rfloor$.

        \STATE Initialize $N_{\text{modified}} = 0$.
        \WHILE{$N_{\text{modified}} < N_{\text{noise}}$}
            \STATE Sample cluster size randomly $s \in \mathbb{R}^{D \times H \times W}, s \sim \text{Uniform}(s_{\text{min}}, s_{\text{max}})$.
            \STATE Sample random spherical coordinates $(\theta, \phi, r')$: \\
                \hspace{0.1cm} $\theta \sim \text{Uniform}(0, 2\pi), \ \phi \sim \text{Uniform}(0, \pi), \ r' \sim \text{Uniform}(0, r)$.
            \STATE Convert to Cartesian coordinates: \\
                \hspace{0.1cm} $x = c_x + r' \sin\phi \cos\theta, \ y = c_y + r' \sin\phi \sin\theta, \ z = c_z + r' \cos\phi$.
            \IF{$(x, y, z)$ is within brain region}

                \FOR{$i, j, k \in [0, s-1]$}
                    \STATE $i' \leftarrow i - \lfloor \frac{s}{2} \rfloor$, $j' \leftarrow j - \lfloor \frac{s}{2} \rfloor$, $k' \leftarrow k - \lfloor \frac{s}{2} \rfloor$
                    \STATE $\mathbf{V}(z+k', y+j', x+i') \leftarrow 0$

                    \STATE $N_{\text{modified}} \leftarrow N_{\text{modified}} + 1$
                    \IF{$N_{\text{modified}} \geq N_{\text{noise}}$}
                        \STATE $\mathbf{\hat{V}} \leftarrow \mathbf{V}$
                        \STATE \textbf{Break}
                    \ENDIF
                \ENDFOR
            \ENDIF
        \ENDWHILE

        \STATE Return $\mathbf{\hat{V}}$
	\end{algorithmic} 
\end{algorithm}

We represent the MRSI volume by $\mathbf{V} \in \mathbb{R}^{D \times H \times W}$, with $D$, $H$, and $W$ being the dimensions in depth, horizontal, and vertical directions. Since the research aims to investigate the effect of the missing-out area on the model's performance, the methodology concentrates on the brain region to make it consistent. First, the algorithm identifies the brain region in each 3D volume $\mathbf{M}$ by masking non-zero pixel values. In this step, the spherical region by two parameters: center coordination $c$ and spherical radius $r$ to make it easier for the blacking-out process is also estimated. Then, the percentage $\eta$ of desired missed-out area $N_{noise}$ compared to the total area of the brain $N_{brain}$ is identified. In our experiment, we consider different portion of the missed areas, including 5\%, 10\%, 15\%, and 20\%, to evaluate the performance of our pipeline. The algorithm randomly generates clusters of varying sizes, $s$, ranging from 1 to 3 voxels to blackout regions of the brain. It also makes sure that the locations of the clusters are random while the clusters are in the brain region. After getting the desired percentage, it was saved for training and testing purposes.

\begin{figure*}[ht!]
    \centering
    \begin{subfigure}[b]{0.85\linewidth}
        \centering
        \includegraphics[width=\linewidth]{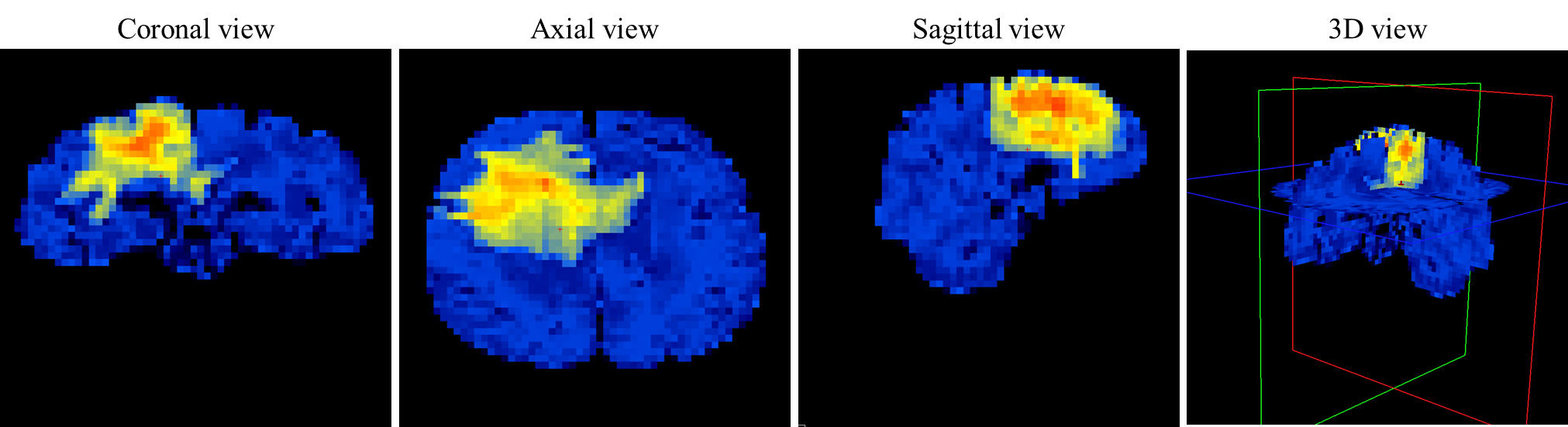}
        \caption{3D MRSI image before preprocessing.}
        \label{fig:before_processing}
    \end{subfigure}
    \begin{subfigure}[b]{0.85\linewidth}
        \centering
        \includegraphics[width=\linewidth]{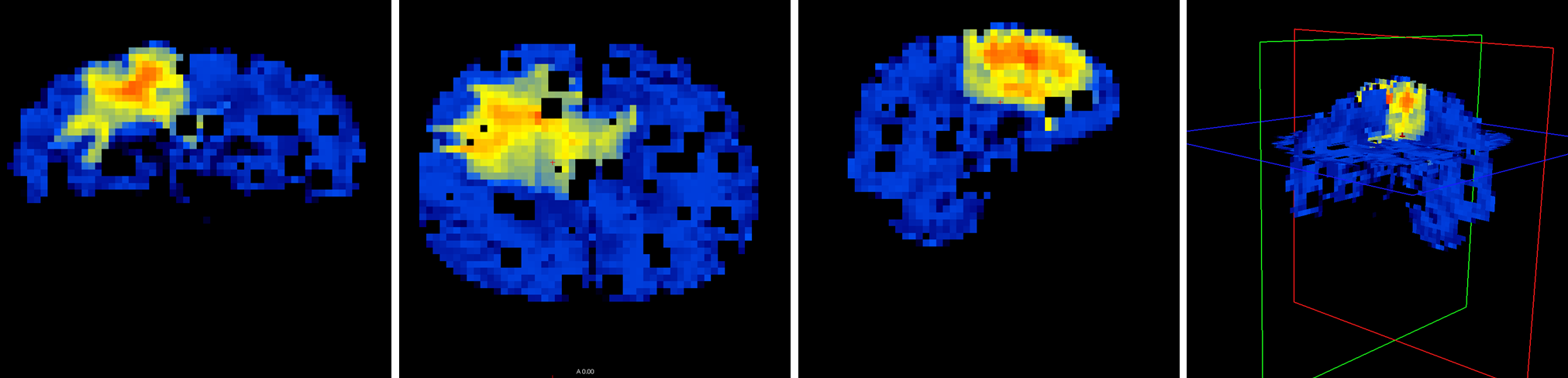}
        \caption{3D MRSI image after preprocessing.}
        \label{fig:after_processing}
    \end{subfigure}
    \caption{Visualization of 3D MRSI metabolic maps before (a) and after (b) preprocessing.}
    \label{fig:dataset}
\end{figure*}

Fig. \ref{fig:dataset} illustrates the original and corrupted versions of a representative 3D MRSI volume with the random removal of 10$\%$ of the voxels. By ensuring consistent dimensions, normalization, and diversity of anomaly sizes, we can closely emulate real-world scenarios involving missing or corrupted MRSI data. This preprocessing is critical for training robust neural networks capable of reconstructing 3D MRSI volumes and filling in missing data accurately. In addition to 3D training, we extract 2D axial slices from the simulated volumes to train a 2D model variant. To minimize the influence of non-informative background regions, we retain only slices where the brain occupies more than 5\% of the area. This ensures a cleaner training signal and improves model convergence in 2D experiments.

\subsection{Deep Learning-Based MRSI Missing Data Estimation}
\label{sec.methodology}

To estimate the missing data in MRSI metabolic maps, we developed a supervised deep-learning framework grounded in the U-Net architecture, implemented in both 2D and 3D formats. The goal of the model is to estimate the complete metabolic map from partially corrupted input volumes, without requiring explicit masks that indicate the location of missing voxels. Our model was designed specifically for volumetric biomedical data with its ability to capture spatial hierarchies and preserve structural features through skip connections.

\begin{figure*}[h]
    \centering
    \includegraphics[width=0.9\linewidth]{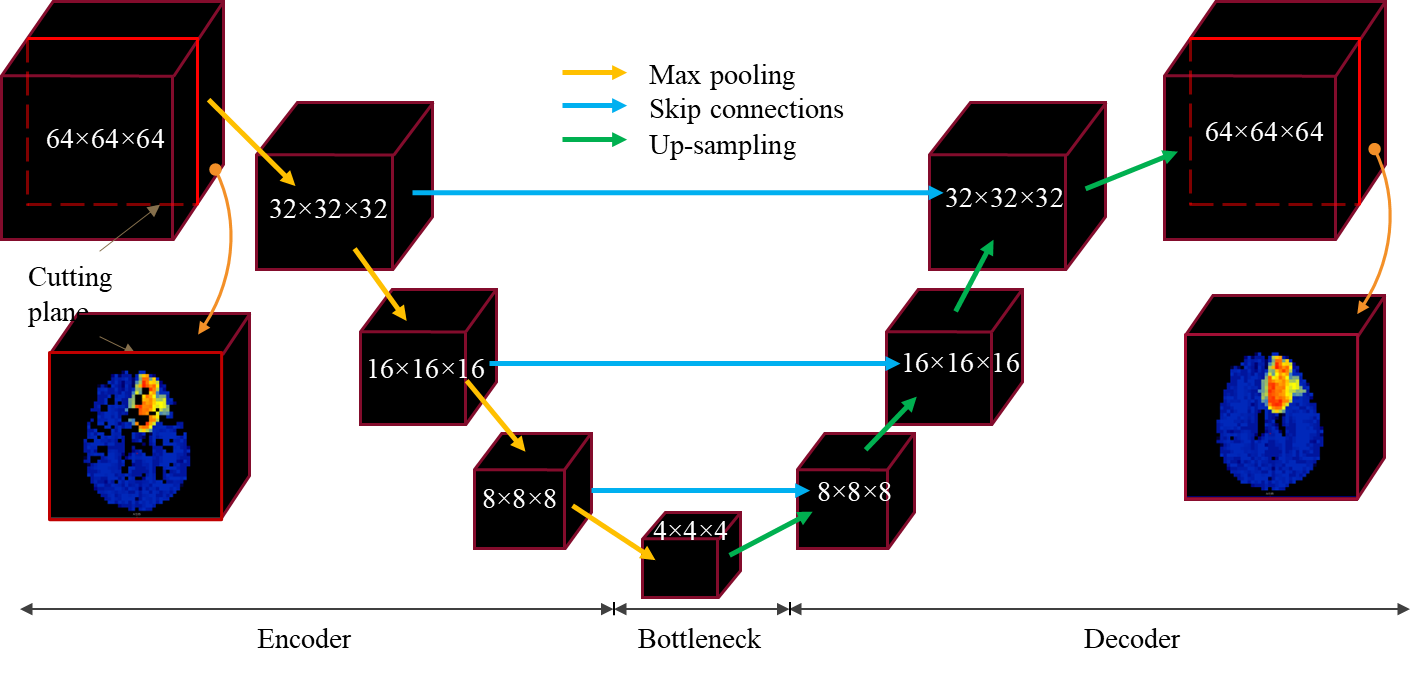}
    \caption{3D deep neural network architecture for MRSI metabolic maps missing data estimation.}
    \label{fig:Unet_model}
\end{figure*}

The proposed architecture is shown in Fig.~\ref{fig:Unet_model}. In our network, the encoder extracts multi-scale features through a series of 3D convolutional layers, each followed by ReLU activation and 3D max-pooling operations. This process progressively reduces the spatial resolution while increasing the depth of the feature representation. The encoder consists of four blocks, where each block reduces the spatial dimensions by half and doubles the number of feature channels, following the sequence: 32, 64, 128, and 256 channels, respectively. In addition, the bottleneck is in the center of the network, which further refines the learned features with 512 channels. The decoder mirrors this process by employing transposed convolutions and up-sampling layers to reconstruct the input image, while ensuring that high-resolution details are recovered. The skip connections between corresponding layers in the encoder and decoder facilitate the direct transfer of spatial information, thereby significantly improving the restoration quality.

Each convolutional layer in the architecture employs a kernel size of $3 \times 3$, a stride of 1, and padding of 1 to maintain spatial dimensions. Batch normalization layers are applied after each convolution to stabilize the training process and accelerate convergence. The output layer employs a single-channel sigmoid activation function to ensure that the reconstructed images remain within the normalized intensity range of $[0,1]$ as mentioned in section \ref{sec.data_processing}.

Recognizing the importance of high-quality prediction, we propose a progressive training strategy to enhance the robustness of the model under varying levels of data degradation. Initially, the model is trained using preprocessed volumes with 5\% missing data, allowing it to adapt to minor imperfections in the input. Once the model demonstrates good performance at this level, the training data is incrementally modified to include 10\% and then 15\% missing data. This gradual increase in the complexity of missing regions encourages the model to learn generalized restoration capabilities, ensuring resilience to larger-scale data loss.

\section{Experimental Evaluation Metrics}
\label{sec.experimentsetup}

The training process leveraged the Adam optimizer with an initial learning rate of $1 \times 10^{-3}$ and a weight decay of $1 \times 10^{-5}$ to prevent overfitting \cite{kingma2014adam}. The learning rate was reduced by a factor of 0.5 every 10 epochs if the validation loss plateaued. The total number of training epochs was set to 500, with early stopping criteria to prevent overfitting when no improvement in validation loss was observed for 10 consecutive epochs. The loss function used for training was a composite loss combining Mean Square Error (MSE) and Structural Similarity Index Measure (SSIM) loss. While MSE ensures pixel-wise accuracy, SSIM emphasizes perceptual quality by comparing luminance, contrast, and structure between the reconstructed and ground-truth images. The composite loss function is defined as:
\begin{equation}
\mathcal{L} = \alpha \cdot \text{MSE} + (1 - \alpha) \cdot (1 - \text{SSIM}),
\end{equation}
where $\alpha$ was empirically set to 0.5 for balanced optimization.

All experiments were conducted on the AI.Panther supercomputer equipped with 4 A100 SXM4 GPUs at the Florida Institute of Technology. The training was performed with a batch size of 32, utilizing the CUDA backend for accelerated computations, and it took approximately 1 hour to train the model. We evaluated model performance using two standard image data missing estimation metrics: MSE and SSIM. Each model was evaluated on both the training and testing datasets, which include synthetic MRSI volumes corrupted with 20\% and 15\% missing voxels for the 2D and 3D datasets, respectively. In addition to these quantitative metrics, we performed qualitative comparisons through the visual inspection of 2D slices and 3D restorations. Visual examples highlighted performance in both healthy and pathological regions, including tumor margins.

\section{Experimental Results}
\label{sec.result}
This section presents a comprehensive evaluation of the proposed framework for estimating missing data in MRSI metabolic maps. In addition to experiments on 3D volumetric reconstruction using simulated data, we extended our study to 2D image restoration tasks. Importantly, we conducted extensive evaluations on 2D real MRSI datasets to demonstrate the model’s practical utility and generalizability in realistic scenarios. However, due to the unavailability of 3D real-world MRSI datasets with known ground truth, we did not perform the evaluation of 3D missing data estimation on real data in this study. 

\subsection{Simulated 2D Missing Data Estimation}

\begin{figure*}[h]
    \centering
    \includegraphics[width=0.9\linewidth]{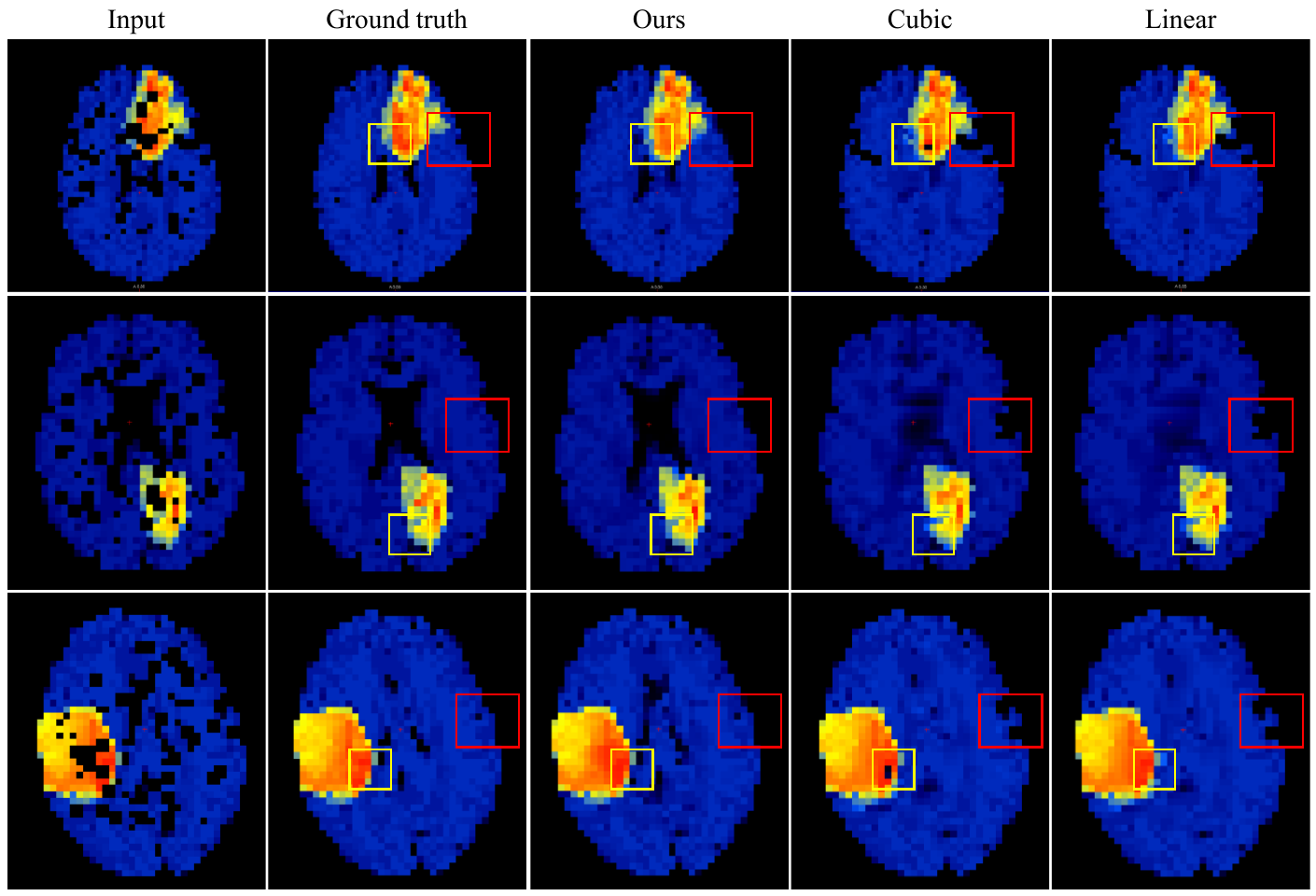}
    \caption{Comparison of 2D simulated MRSI metabolic map restoration results on 20\% missing data using the proposed deep learning method and traditional interpolation techniques. Beyond the differences in the ventricular regions at the center of the images, the yellow boxes highlight reconstruction discrepancies in tumor-affected areas, while the red boxes indicate notable differences in healthy brain regions. These areas illustrate the superior structural preservation and detail recovery achieved by the proposed approach.}
    \label{fig:image_comparison}
\end{figure*}

Table~\ref{tab.2D_result} presents the quantitative metrics, including MSE and SSIM between our method and interpolation methods such as cubic and linear interpolation. The experiments were conducted using synthetic data with 20\% of the brain region removed. In addition, we also included the progressive training strategy compared to the direct training strategy for our approach. As shown in Table~\ref{tab.2D_result}, our method significantly outperforms conventional interpolation techniques, achieving an MSE of 0.002 on the test set and an SSIM of 0.97. In terms of our training approaches, the progressive training strategy performs slightly better than the direct training strategy. In contrast, cubic interpolation results in an MSE of 0.005 and an SSIM of 0.94, while linear interpolation yields an MSE of 0.004 and an SSIM of 0.95. These results demonstrate the superior restoration fidelity of our deep learning-based approach.  Fig.~\ref{fig:image_comparison} provides a visual comparison among the various methods, where the red and yellow boxes highlight key areas of differences in the normal region and tumor region, respectively. The proposed method exhibits better preservation of structural details and smoother transitions, particularly in the tumor-affected regions. Notably, the proposed method correctly preserved the ventricular region without treating it as missing data, whereas conventional interpolation methods erroneously interpolated across this area, leading to anatomically inaccurate reconstructions.

\begin{table}[h]
\centering\small
\caption{Quantitative evaluation of the performance of our proposed method compared to interpolation methods when 20\% of the brain region is removed.}
\label{tab.2D_result}
\begin{tabular}{l|cc|cc}
\toprule
\textbf{Methodology} & \multicolumn{2}{c|}{\textbf{Training Data}} & \multicolumn{2}{c}{\textbf{Test Data}} \\
\cmidrule(lr){2-3} \cmidrule(lr){4-5}
 & MSE &  SSIM  & MSE & SSIM \\
\midrule 
Ours (Progressive training)& 0.001 & 0.98 &  0.002 &  0.97  \\
Ours (Direct training) & 0.002   & 0.96 &  0.002 & 0.96  \\
\midrule
Cubic interpolation & 0.005 & 0.93 & 0.005  & 0.94\\
Linear  & 0.004& 0.94 & 0.004 &  0.95\\
\bottomrule
\end{tabular}
\end{table}

\subsection{Simulated 3D Missing Data Estimation}

\begin{figure}[h]
    \centering
    \begin{subfigure}[b]{.95\linewidth}
        \centering
        \includegraphics[width=\linewidth]{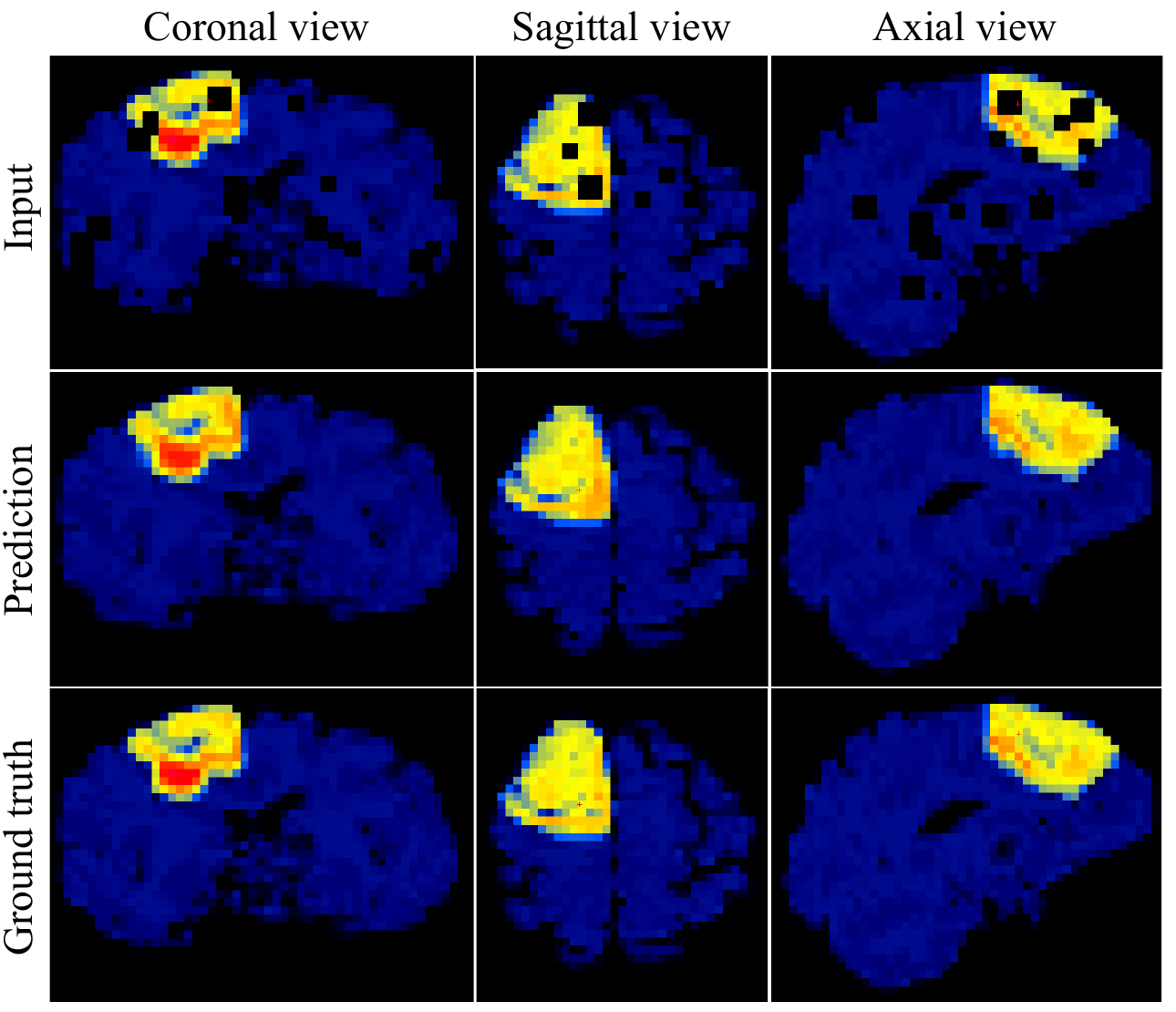}
        \caption{Missing data estimation of sample 1.}
        \label{fig:sample1}
    \end{subfigure}
    \hfill
    \begin{subfigure}[b]{.95\linewidth}
        \centering
        \includegraphics[width=\linewidth]{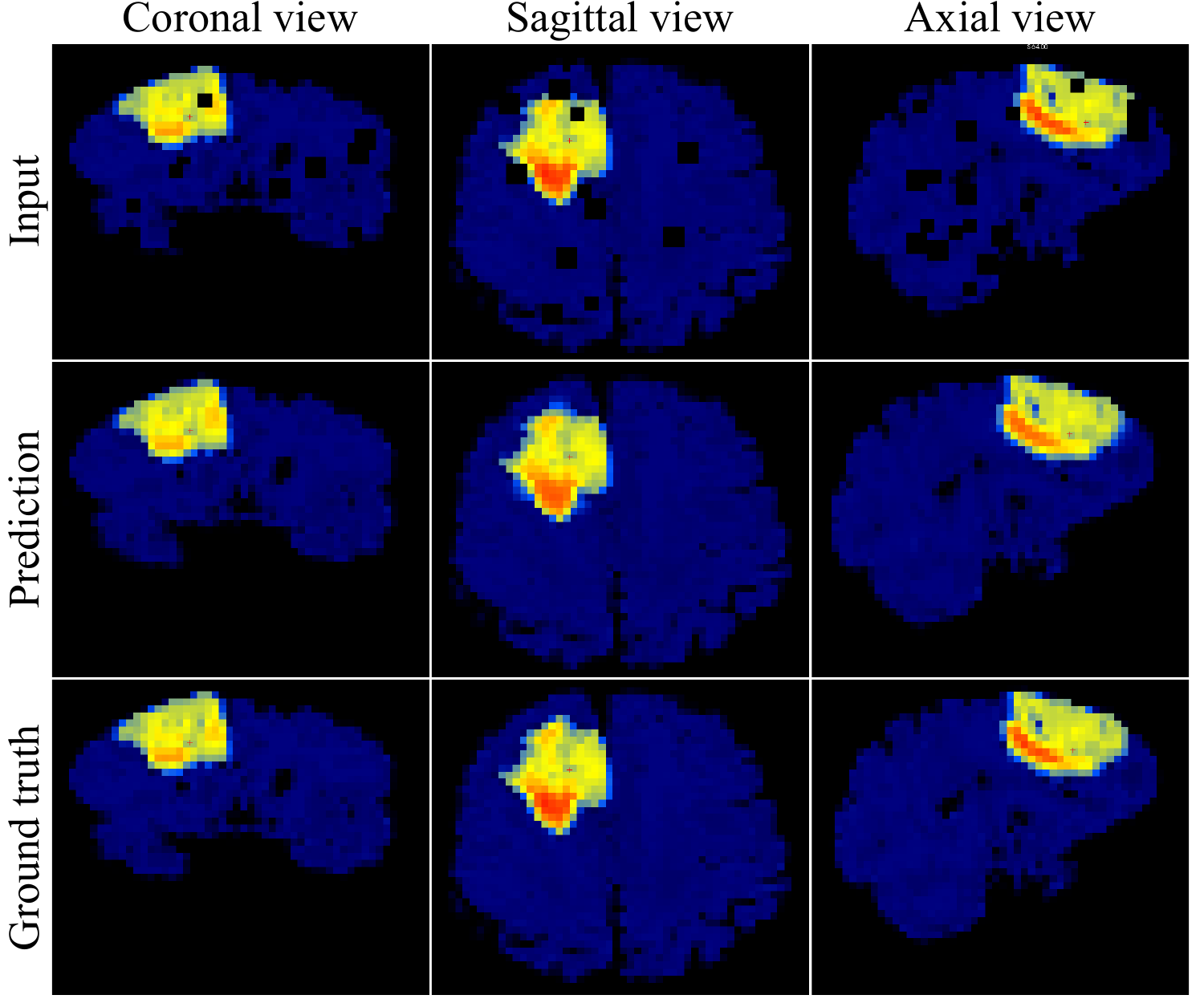}
        \caption{Missing data estimation of sample 2.}
        \label{fig:sample2}
    \end{subfigure}
    \caption{Missing data estimation by the proposed model on simulated 3D metabolic maps with 15\% missing data: Sample 1 (a)  and Sample 2 (b). In each sample, the coronal, sagittal, and axial views are shown from left to right, respectively, while the input, the model's prediction, and the ground truth are shown from top to bottom.}
    \label{fig:3D_prediction}
\end{figure}

The quantitative evaluation results of the proposed methodology on 3D volumetric data missing estimation were summarized in Table~\ref{tab.3D_result}, which demonstrate the efficacy of the proposed method in estimating missing voxels with superior performance compared to traditional interpolation techniques. Specifically, our approach achieves an MSE of 0.001 on both the training and test sets, significantly outperforming cubic and linear interpolation methods. The SSIM score of 0.98 for both sets further highlights the model's capability to reconstruct volumetric data with high fidelity and structural coherence.

Based on the information from Table~\ref{tab.2D_result} and Table~\ref{tab.3D_result}, we can conclude that the MSE on the 3D dataset is similar to that on the 2D dataset. This is due to the 3D dataset containing a larger proportion of voxels with smaller values, which contribute less variation and result in smaller restoration errors, thereby yielding lower MSE values. Similarly, the SSIM scores are comparable across both datasets, indicating that the model consistently preserves structural details regardless of differences in voxel intensity distributions. For comparison, cubic interpolation achieves an MSE of 0.002 and a lower SSIM of 0.93, while linear interpolation yields an MSE of 0.008 and an SSIM of 0.05. These results highlight the superiority of our approach in capturing fine details and preserving the structural integrity of 3D volumetric data. Intuitively, Fig.~\ref{fig:3D_prediction} provides a qualitative evaluation, showcasing the reconstructed 3D volumetric structures. As illustrated, our method effectively estimates the missing regions while accurately preserving the existing structures, closely matching the ground truth.

\begin{table}[h]
\centering\small
\caption{Evaluation of the performance of our proposed method compared to interpolation methods on the 3D dataset, with 15\% of the brain region removed.}
\label{tab.3D_result}
\begin{tabular}{l|cc|cc}
\toprule
\textbf{Methodology} & \multicolumn{2}{c|}{\textbf{Training Data}} & \multicolumn{2}{c}{\textbf{Test Data}} \\
\cmidrule(lr){2-3} \cmidrule(lr){4-5}
 & MSE &  SSIM  & MSE  & SSIM \\
\midrule 
Ours (Progressive training) & 0.001 & 0.98 &  0.001 &  0.98  \\
Ours (Direct training) & 0.002  & 0.97 & 0.002  &   0.97 \\
\midrule
Cubic interpolation & 0.002 & 0.93 & 0.002 & 0.93\\
Linear  & 0.008 & 0.95 & 0.009&  0.95\\
\bottomrule
\end{tabular}
\end{table}

\subsection{Inference on Real Datasets}

\begin{figure}[h]
    \centering
    \includegraphics[width=1\linewidth]{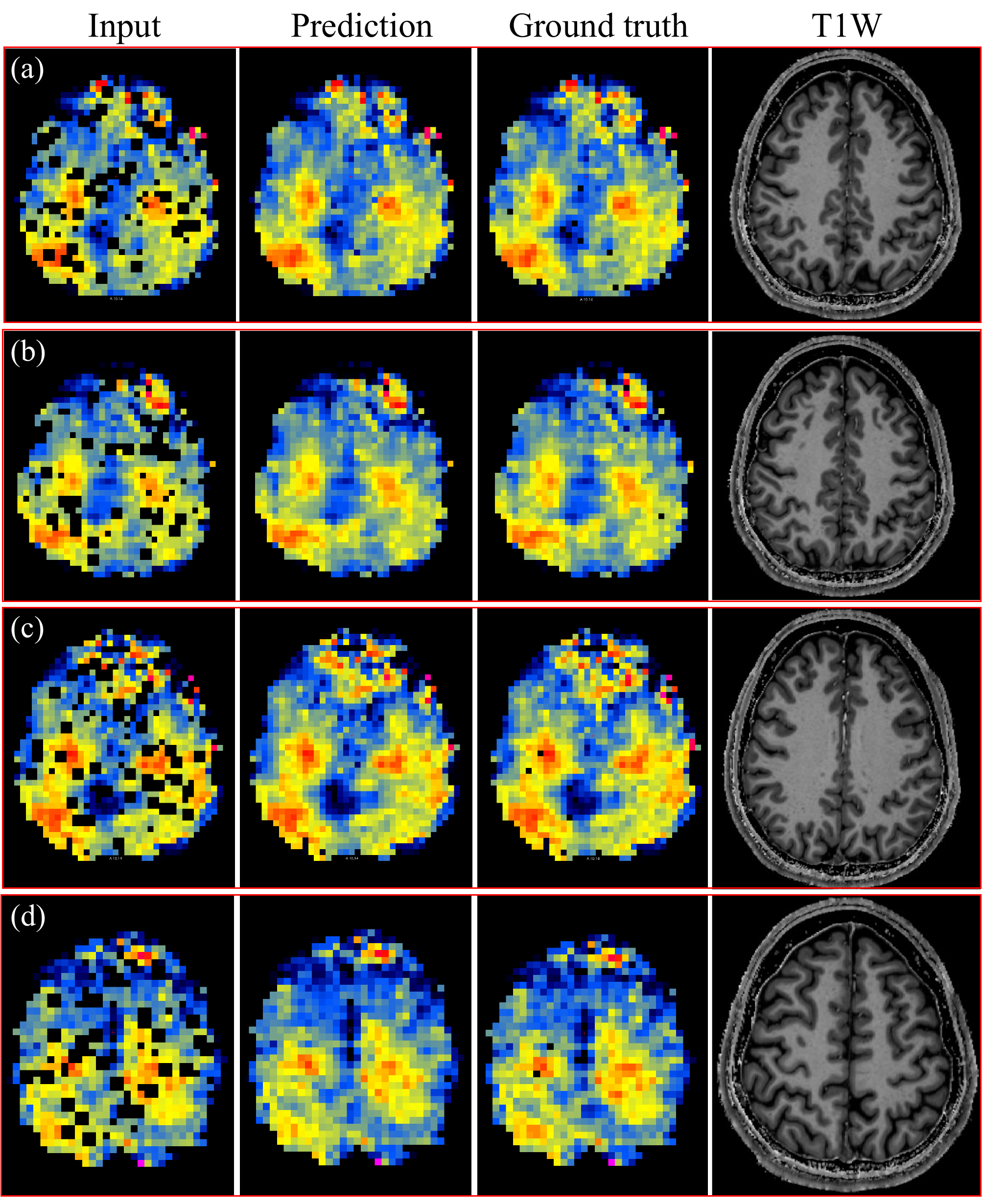}
    \caption{Restoration results for 15\% missing data estimation in 2D N-acetylaspartate (NAA) metabolic maps from real healthy subject data. The proposed model was tested on four representative samples (a–d). For each sample, the first column displays the input image with simulated missing voxels, the second column shows the corresponding reconstruction generated by the model, and the third column presents the ground truth metabolic maps for reference.
    }
    \label{fig:real_patient}
\end{figure}

To evaluate the generalizability of our approach beyond the simulated dataset, we tested it on real-world datasets. Notably, no real patient data was used during the training process. Fig. \ref{fig:real_patient} shows the results of the estimation of missing data in real healthy datasets provided in \cite{hangel2021inter}. Specifically, we evaluated the model on four samples: a, b, c, and d, in which each with a size of $64 \times 64$. In each case, the first column shows the image with missing data to be restored, the second column presents the model's prediction, and the third column displays the ground truth. For the images with missing data, we applied the pipeline described in Section~\ref{sec.data_processing} to generate inputs with 15\% of the brain region removed. As a result, despite some outliers in the dataset, our method successfully reconstructed the missing regions, demonstrating its ability to preserve anatomical structures and produce high-fidelity restorations.

\begin{figure}[ht]
    \centering
    \includegraphics[width=1\linewidth]{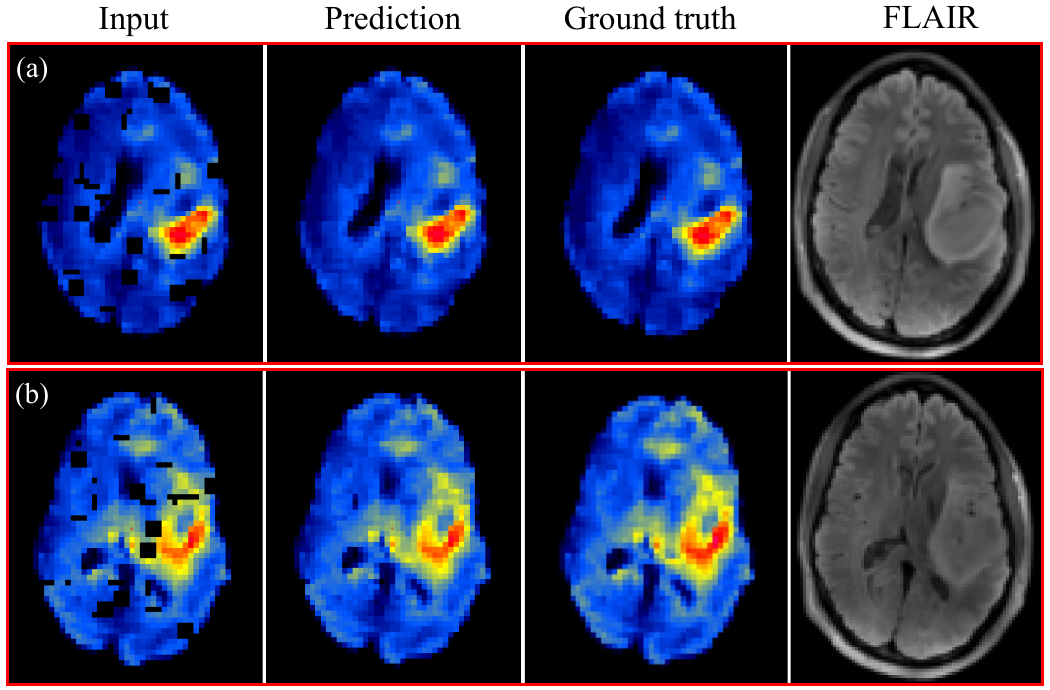}
    \caption{Restoration results for 15\% missing data estimation in 2D total choline (tCho) metabolic maps from real patient data. The input data, with a size of $64 \times 64$, were generated by randomly removing 15\% of the brain regions.}
    \label{fig:real_patient_origres}
\end{figure}

\begin{figure}[ht]
    \centering
    \includegraphics[width=0.985\linewidth]{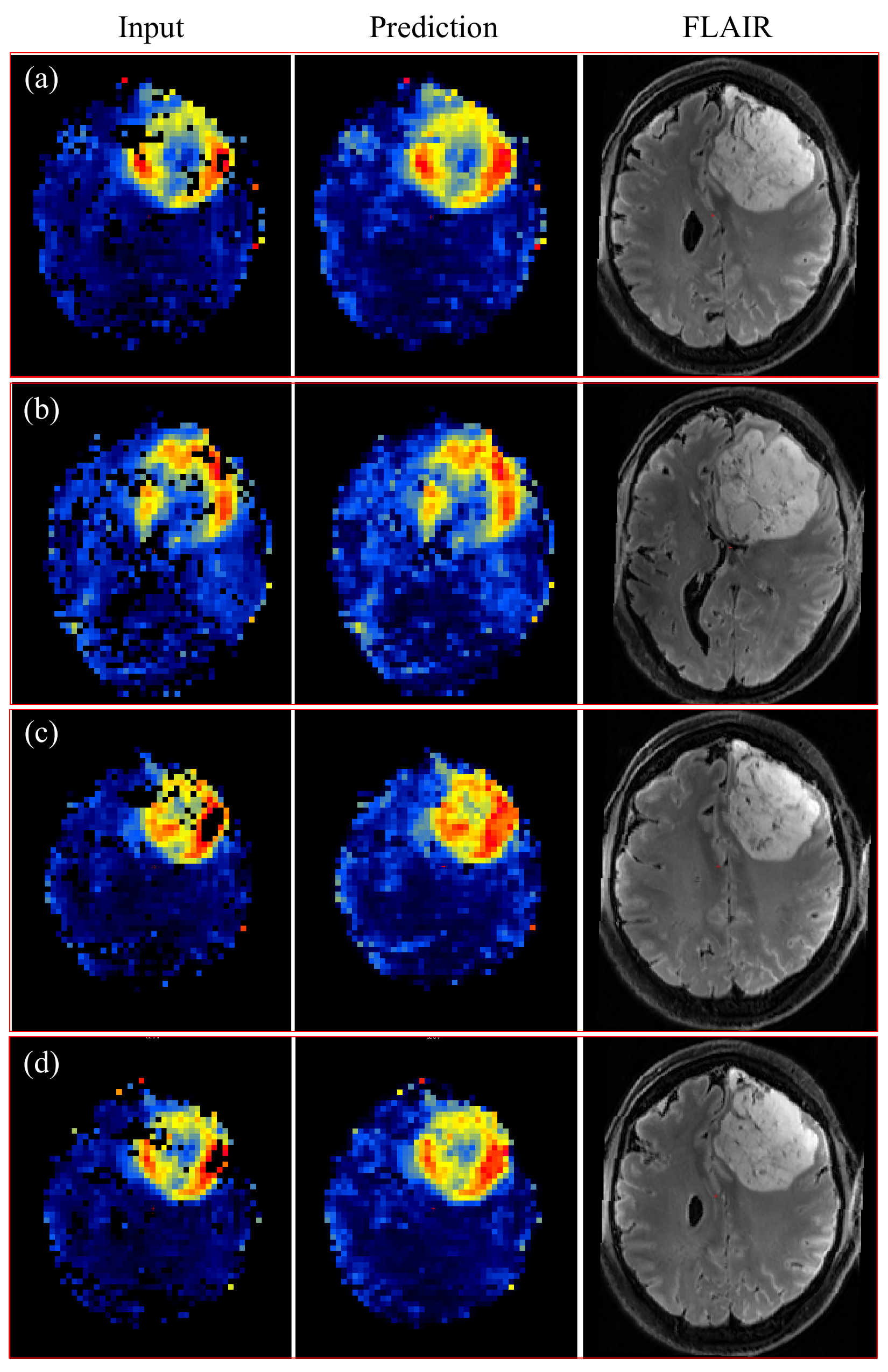}
    \caption{Inference outcomes for estimating missing data in 2D total Choline (tCho) metabolic maps from real patient data.}
    \label{fig:real_patient_map1}
\end{figure}

Similarly to Fig. \ref{fig:real_patient}, we extend our experiments to 2D total choline (tCho) metabolic maps of size $64 \times 64$ from a real patient dataset. The restoration results are presented in Fig. \ref{fig:real_patient_origres}. As in the previous experiments on simulated data, we employed the method proposed in section \ref{sec.data_processing} to generate a dataset with 15\% missing areas and then applied our model to restore the missing regions. As illustrated, our model accurately predicts the missing areas for the input images. We then assess the generalization capability of our method using a real-world dataset \cite{hangel2021inter} that contains naturally occurring missing regions but lacks ground truth, as illustrated in  Fig. \ref{fig:real_patient_map1}. The experiment was carried out on four tCho metabolic maps, each with an image size of $64 \times 64$. The results indicate that our method effectively predicts missing regions in both tumor-affected and normal brain areas. These findings further demonstrate the potential of our approach for reconstructing missing data in MRSI, underscoring its value in enhancing the clarity and reliability of medical imaging.

\section{Discussion and Conclusion}
\label{sec.discussion}

In this study, we propose a novel deep learning-based approach for missing data estimation for MRSI. To the best of our knowledge, this is the first work to address MRSI restoration using a mask-free strategy. Despite being trained on a limited set of simulated data, our model demonstrates strong generalization to both synthetic and real-world datasets. It significantly outperforms traditional interpolation techniques, achieving an MSE of 0.002 and an SSIM of 0.97 on 2D test data, and an MSE of 0.001 with an SSIM of 0.98 on 3D volumetric data, while effectively preserving fine structural and spatial details. Evaluation on real patient data further confirms the robustness of the proposed method, highlighting its potential for reliable metabolic image reconstruction in clinical scenarios.

Our findings represent a notable step forward in medical imaging, particularly for diagnosis, treatment planning, and monitoring. Unlike conventional MRI, MRSI reveals underlying metabolic and biochemical changes. Accurate restoration of MRSI enables high-resolution 3D visualization of metabolite distributions, aiding in the early detection of conditions such as as brain tumors, epilepsy, and neurodegenerative diseases. This also supports integration with other imaging modalities and improves surgical planning and disease tracking. Although our method performs well, it currently depends on simulated training data. Future efforts will focus on real-time inference and clinical integration.

\section*{Acknowledgments}
\small{This work was partially supported by the USDA NIFA grant ${\#}$2022-67021-38911 and NSF grant ${\#}$2138206 \& ${\#}$2410678.}

\bibliographystyle{IEEEtran}
\bibliography{reference}

\end{document}